\documentclass[twocolumn,aps,prd,showpacs,nofootinbib,superscriptaddress,floatfix,showkeys]{revtex4-2}

\usepackage{amsfonts}
\usepackage{amsmath}
\usepackage{amssymb} 
\usepackage{bm} % For bold math (esp of lower greek letters)
\usepackage{color}
\usepackage{enumerate}
\usepackage{graphicx}  % Needs to come after graphicx
\usepackage{dcolumn}
\usepackage{hyperref}
\usepackage{lineno}
\hypersetup{
	colorlinks=true,       % false: boxed links; true: colored links
	linkcolor=blue,          % color of internal links
	citecolor=magenta,        % color of links to bibliography
	filecolor=cyan,      % color of file links
	urlcolor=magenta           % color of external links
}

\labelformat{section}{#1} 
\labelformat{subsection}{Section #1} 
\labelformat{figure}{#1} 
\labelformat{subfigure}{Fig.~\thefigure#1} 
\labelformat{table}{Tab.~#1} 

\def\be{\begin{equation}}
	\def\ee{\end{equation}}
\def\ben{\begin{eqnarray}}
	\def\een{\end{eqnarray}}

\begin{document}
\title{Gravitational Waves from Primordial Black Holes: Connecting Low-Frequency Scalar-Induced Signatures to High-Frequency Binary Mergers}
\author{Ashu Kushwaha} 
\email{kushwaha.a.ce1b@m.isct.ac.jp}
\affiliation{Department of Physics, Institute of Science Tokyo, 2-12-1 Ookayama, Meguro-ku,
	Tokyo 152-8551, Japan}
%
%	\author{}
%	\email{}
%	\affiliation{}
	%
	%
   % \date{\today}
\begin{abstract}
Formation of primordial black holes (PBHs) requires a significant enhancement of curvature perturbations. This enhancement generates a twofold gravitational wave (GW) signature: a \emph{low-frequency} stochastic background of scalar-induced GWs (SIGWs) and a distinct \emph{high-frequency} signal from subsequent PBH binary mergers. Assuming a monochromatic PBH mass function, we use PBH abundance constraints on the primordial curvature power spectrum to evaluate the stochastic SIGW background. 
We also compute the stochastic GW background from mergers of the corresponding PBH binaries, incorporating merger-rate suppression effects to obtain realistic estimates. Furthermore, we derive a model-independent correspondence between the characteristic frequencies of these two signals. 
This unified framework links these otherwise distinct GW channels, enabling the same primordial fluctuations to be probed across widely separated frequency bands.

\end{abstract}
\maketitle
%\newpage

%
\textbf{Introduction.}
Primordial black holes (PBHs) can form in the early Universe from the gravitational collapse of large-amplitude density fluctuations~\cite{Zeldovich:1967lct,1971-Hawking-MNRAS,Carr:1974nx,Chapline:1975ojl,Khlopov:2008qy,2016-Carr.Kuhnel.Sandstad-PRD,2018-Sasaki.etal-CQG,2021-Carr.etal-Rept.Prog.Phys,2020-Green.Kavanagh-JPhyG,Escriva:2022duf,Shankaranarayanan:2026hnn}. Specifically, when enhanced primordial curvature perturbations generated by nontrivial physics during inflation (see Refs.~\cite{Garcia-Bellido:1996mdl,Yokoyama:1998pt,Kohri:2007qn,Kawasaki:2012wr,Garcia-Bellido:2017mdw,Inomata:2017okj,Carr:2018nkm,Byrnes:2018txb,Inomata:2018cht,Bhaumik:2019tvl,Braglia:2020eai,Ragavendra:2020sop,Cai:2023uhc}), re-enter the Hubble horizon during radiation domination (RD), they generate overdense regions that collapse into PBHs if their amplitude exceeds a critical threshold. Therefore, PBH formation provides a sensitive probe of the primordial power spectrum in regimes inaccessible to cosmic microwave background (CMB) and large-scale structure observations. 

The enhanced primordial curvature perturbations responsible for PBH formation inevitably generate a stochastic background of scalar-induced gravitational waves (SIGWs) at second order when they re-enter the Hubble horizon during the RD era~\cite{2018-Caprini.Figueroa-CQG,Inomata:2018epa,2021-Domenech-Universe-GW-Review}. Since gravitational waves interact only weakly with matter, they preserve information about their production mechanism over cosmological timescales~\cite{1999-Maggiore-PhyRept,2004-Bisnovatyi.Kogan-CQG,2009-Sathyaprakash.Schutz-LivRevRel,Book-Maggiore-Vol2}. Consequently, SIGWs provide a complementary probe of the small-scale primordial fluctuations responsible for PBH formation~\cite{Baumann:2007zm,Saito:2008jc}. Various present and future GW observatories aim to explore these singlas over a very broad frequency range, including pulsar timing arrays (PTAs)~\cite{NANOGrav:2015aud,Shannon:2015ect,EPTA:2015qep} LIGO-Virgo-KAGRA~\cite{LVK:2013rdx}, LISA~\cite{Barausse:2020rsu}, DECIGO~\cite{Kawamura:2020pcg}, the Einstein Telescope (ET)~\cite{ET:2019dnz}, and Cosmic Explorer (CE)~\cite{2018-Caprini.Figueroa-CQG,2021-Domenech-Universe-GW-Review}.

If PBHs form binaries, their mergers generate a second \emph{distinct} GW signal characterized by the total binary mass, $M_t = m_1 + m_2$~\cite{Mandic:2016lcn,Raidal:2017mfl,Raidal:2018bbj,Liu:2021jnw,Wang:2019kaf,Gow:2019pok,Vaskonen:2019jpv,Hall:2020daa,Pujolas:2021yaw,Bavera:2021wmw,Aggarwal:2025noe,Yu:2026vey}, a scenario widely investigated to explain the binary black hole merger event observed by the LIGO-Virgo collaboration~\cite{LIGOScientific:2016aoc,Bird:2016dcv,Clesse:2016vqa,Sasaki:2016jop}. The inspiral emission is nearly maximal at the innermost stable circular orbit (ISCO), which marks the onset of the merger phase at frequency $f_{\rm ISCO}$. Since the PBH mass is approximately equal to the horizon mass at their formation epoch, it is directly tied to the characteristic peak scale (comoving wavenumber, $k_p$) of the enhanced primordial curvature power spectrum. % 
Therefore, this establishes an approximate analytical connection between the peak frequency of the SIGW background, $k_p \simeq 2\pi f_{\rm SIGW}$, and the characteristic merger frequency, yielding $f_{\rm ISCO} \propto f_{\rm SIGW}^2$. Remarkably, this relation demonstrate that the initial primordial enhancement can produce a twofold observable GW signature: a low-frequency stochastic SIGW background and a corresponding high-frequency PBH merger signal.
For example, the PTA frequency band for SIGWs maps onto the Advanced LIGO (aLIGO) and ET bands for PBH mergers~\cite{Cang:2023ysz}. These multifrequency scenarios have potential implications for probing the high-frequency GW background, whose detection remains at an early stage.~\cite{Aggarwal:2025noe}. 

In this \textit{Letter}, we establish the connection between these two distinct GW observables, proving that they can act as complementary probes of the same primordial enhancement.
Assuming a narrow-peaked enhancement in the primordial power spectrum, we apply recent constraints on its amplitude derived from the PBH abundance~\cite{Kushwaha:2026msi} to systematically compute the resulting low-frequency stochastic SIGW background.
We present the first consistent comparison of the SIGW spectra inferred using the spherical-collapse approximation and its more realistic ellipsoidal refinement.
We then evaluate the high-frequency GW signals generated by mergers of the same PBHs responsible for the SIGW signal, explicitly comparing the idealized unsuppressed case with the more realistic scenario including merger-rate suppression effects.
By mapping the peak frequencies of both the SIGW and merger signals, we derive a model-independent relation that connects these widely separated spectral bands. This unified framework thus provides a novel, comprehensive approach to assessing these complementary GW channels and refining PBH source modeling.

\textbf{PBH formation. }
In the Press-Schechter formalism, assuming Gaussian density perturbations, the PBH mass fraction $\beta$ at formation is defined as~\cite{1974-Press-Schechter-APJ,Kushwaha:2025zpz,2018-Sasaki.etal-CQG}
\begin{align}\label{beta-ps}
    \beta (M) = \gamma\int_{\delta_{\rm th}}^{\delta_{\rm max}} d\delta ~  \frac{1}{\sqrt{2\pi} \, \sigma} \exp{\left(-\frac{\delta^2}{2\sigma^2}\right)}, 
\end{align}
where $\delta$ is the smoothed density contrast,
$\delta_{\rm max}=4/3$, and $\gamma \simeq 0.2$ parameterizes the efficiency of gravitational collapse. The horizon mass $M$ corresponds to the comoving scale $k_m^{-1}$ at Hubble horizon reentry.
Although the primordial curvature perturbation $\zeta$ is assumed to follow Gaussian statistics, the non-linear relation between $\zeta$ and $\delta$ makes the density contrast non-Gaussian. To incorporate these non-linearities while retaining a tractable Gaussian probability distribution function, we adopt the mapping framework from Refs.~\cite{Young:2019yug,Kushwaha:2026msi}. The non-linear density contrast $\delta_{\rm NL}$ maps to its linear counterpart $\delta_L$ via $\delta_c \equiv \delta_{\rm NL} = \delta_L - (3/8)\delta_L^2$. Solving for the linear component yields an effective threshold density $\delta_{\rm th} = \frac{4}{3} ( 1 - \sqrt{1 - 3\delta_{c}/2} )$, where $\delta_c = 0.45$ is the critical threshold. For notational simplicity, we suppress the subscript $L$ hereafter and evaluate Eq.~\eqref{beta-ps} entirely in terms of the linear component.

The variance $\sigma^2$ of the density contrast is related to the primordial curvature power spectrum $\mathcal{P}_{\zeta}(k)$ by
\begin{align*}%\label{sigma-l}
    \sigma^2 (r_m)= \frac{16}{81} \int_0^{\infty} d\ln{k} \, (kr_m)^4 \tilde{W}^2(k,r_m) T^2 (k,r_m) \mathcal{P}_{\zeta} (k),
\end{align*}
where $r_m = 1/k_m$ is the comoving smoothing scale and $k$ is the comoving wavenumber. For a real-space top-hat window function, $\tilde{W}(k,r_m) = 3 [ \sin{x} - x \cos{x}]/x^3$, while the RD era transfer function is $T(k,r_m) = 3 [ \sin(x/\sqrt{3}) - (x/\sqrt{3})\cos(x/\sqrt{3})]/(x/\sqrt{3})^3$ with $x = kr_m$. Non-spherical effects during collapse are incorporated by defining an ellipsoidal threshold as $\delta_{\rm ec} \simeq \delta_{\rm th} [ 1+ 9\sigma / (\sqrt{10\pi}\delta_{\rm th} ) ]$~\cite{Kuhnel:2016exn,Kushwaha:2026msi}, which replaces the spherical collapse threshold, $\delta_{\rm th}$, in Eq.~\eqref{beta-ps}.
We consider a narrow-peaked log-normal type primordial power spectrum $\mathcal{P}_{\zeta} (k) = \frac{\mathcal{A}_p}{\sqrt{2\pi}\Delta} \exp\left( - \frac{\ln^2(k/k_p)}{2\Delta^2}\right)$, 
%
%\begin{align}\label{lognormal}
%    \mathcal{P}_{\zeta} (k) = \frac{\mathcal{A}_p}{\sqrt{2\pi}\Delta} \exp\left( - \frac{\ln^2(k/k_p)}{2\Delta^2}\right),
%\end{align}
where $\mathcal{A}_p$, $k_p$, and $\Delta$ denote the amplitude, position, and width of the peak, respectively. Following Ref.~\cite{Kushwaha:2026msi}, we adopt $\Delta=0.1$, which yields a monochromatic PBH mass function, such that all PBHs form with the same mass. This assumption is adopted throughout this work. As shown in Ref.~\cite{Kushwaha:2026msi}, the peak theory and the Press-Schechter formalism yield identical constraints on $\mathcal{A}_p$, implying that our estimates of the stochastic GW energy density are insensitive to the choice of PBH formation framework.

For a sufficiently narrow peak considered here, the PBH formation scale $k_m$ is related to the peak scale as $k_m \simeq k_p$. The correspondence is approximate because the peak of the smoothed density perturbations is shifted relative to that of the primordial curvature power spectrum, with the magnitude of the shift depending on the choice of window function (see Ref.~\cite{Wang:2019kaf}). Since this shift is small for narrow peaks, we identify the two scales and express the PBH mass as~\cite{2018-Sasaki.etal-CQG}
\begin{align}\label{mpbh-k-relation}
    M_{\rm PBH} \simeq 30 M_{\odot} \left( \frac{g_{*}}{10.75} \right)^{-1/6} \left(\frac{k_p}{2.9\times 10^5 \, \mathrm{Mpc}^{-1}} \right)^{-2},
\end{align}
where $g_{*}$ is the number of relativistic degrees of freedom at the epoch of PBH formation.

\textbf{Scalar-induced gravitational wave background. }
In this section, we briefly review the production of SIGWs, which arise at second order from primordial scalar perturbations (see, e.g., Refs.~\cite{Kohri:2018awv,2018-Caprini.Figueroa-CQG,2021-Domenech-Universe-GW-Review,Pi:2020otn}). 
For the curvature perturbations following a Gaussian distribution,
the GWs induced during RD era in conformal Newtonian gauge can be given by~\cite{Kohri:2018awv,Inomata:2018epa,2021-Domenech-Universe-GW-Review}
    \begin{align}\label{omega_GW-rd}
        \Omega_{\rm GW} (\eta,k) = \frac{1}{24} \left( \frac{k}{a(\eta) H(\eta)} \right)^2 \overline{\mathcal{P}_h ~(\eta,k)} ~ ,
    \end{align}
 where $a(\eta)$, $H(\eta)$, and $\eta$ denote the scale factor, Hubble parameter, and conformal time, respectively. %
 The tensor power spectrum in the above equation is given by~\cite{Kohri:2018awv,Inomata:2018epa,2021-Domenech-Universe-GW-Review}
\begin{align}
     \overline{\mathcal{P}_h (\eta,k)} \simeq &4 \int_0^{\infty} dv \int_{|1-v|}^{1+v} du \left( \frac{4v^2 - (1+v^2-u^2)^2}{4vu} \right)^2 \nonumber\\ 
     & \qquad \overline{I^2(v,u,k\eta)}  \, \,\mathcal{P}_{\zeta} (kv) \mathcal{P}_\zeta (ku)~~,
 \end{align}
where the integration kernel $ \overline{I^2(v,u,k\eta)} $ in the subhorizon limit i.e., $x\equiv k\eta \rightarrow\infty$ is~\cite{Kohri:2018awv,Inomata:2018epa,2021-Domenech-Universe-GW-Review}
\begin{align*}
     \overline{I^2(v,u,x)}  %
     &= \frac{1}{2} \left( \frac{ 3(u^2+v^2-3)^2}{ 4u^3v^3 x} \right)^2 \left[\left( - \frac{4uv}{u^2+v^2-3} \right. \right. \nonumber \\
     & \left.\left. + \log \left| \frac{3-(u+v)^2}{3-(u-v)^2} \right|\right)^2 + \pi^2  \Theta (u+v-\sqrt{3}) \right]
\end{align*}
where $\Theta$ denotes the Heaviside step function. 

Since $\overline{I^2}\propto (k\eta)^{-2}$ in the subhorizon limit, this dependence is exactly canceled by the prefactor $(k/aH)^2=k^2\eta^2$ in Eq.\eqref{omega_GW-rd} during the RD era, implying that the production of induced GWs is dominated around horizon reentry. Consequently, the induced GW energy density becomes time-independent, indicating that GW production effectively ceases after horizon reentry.
We therefore define $\eta_c$ as the time after which $\Omega_{\rm GW}$ becomes time-independent, which is approximately equal to the horizon-reentry time.
For the modes considered here, $\eta_c<\eta_{\rm eq}$, where $\eta_{\rm eq}$ denotes the matter-radiation equality epoch. Accounting for the subsequent evolution through matter-radiation equality and the change in the effective relativistic degrees of freedom, the present-day GW energy density is given by~\cite{Inomata:2018epa,2021-Domenech-Universe-GW-Review}
\begin{align}\label{gw-present}
    \Omega_{\rm GW,0} (k) = 0.83 \left( \frac{g_c}{10.75}\right)^{-1/3} \Omega_{\gamma,0} ~ \Omega_{GW} (\eta_c,k)
\end{align}
where $\Omega_{\gamma,0} = \rho_{\gamma,0}/\rho_{\rm cr}=2.4\times 10^{-5} \, h^{-2}$ is the present-day radiation energy density parameter, $g_c$ is the effective relativistic degrees of freedom at the epoch of reentry $\eta_c$.
\begin{figure}[t!]
\centering
%\hspace{.2cm}
\includegraphics[height=2.6in,width=3.4in]{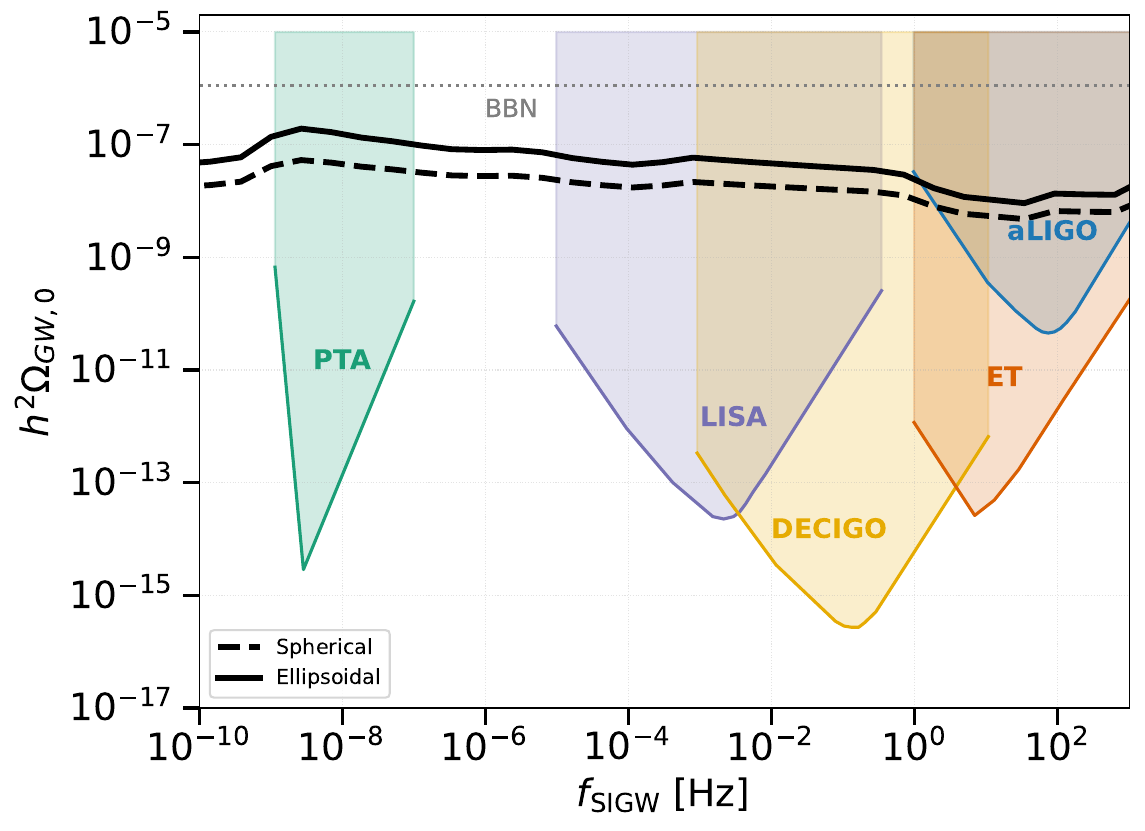}
\caption{Present-day SIGW energy density obtained using the constraints on $\mathcal{A}_p$ from the PBH abundance in Ref.~\cite{Kushwaha:2026msi} for a monochromatic PBH mass function ($\Delta=0.1$). The black solid (dashed) curves correspond to the ellipsoidal (spherical) collapse scenario. Sensitivity curves for PTA, LISA, DECIGO, ET, and Advanced LIGO (aLIGO) are also shown. The gray dotted line denotes the BBN bound~\cite{2018-Caprini.Figueroa-CQG,Aggarwal:2025noe}. The SIGW frequency is approximately related to the peak of the primordial curvature power spectrum as $k_p\simeq 2\pi \, f_{\rm SIGW}$.}
\label{fig:sigw-constraints}
\end{figure}

Equations~\eqref{omega_GW-rd} and \eqref{gw-present} allow us to compute the present-day GW spectrum directly from a given primordial curvature power spectrum $\mathcal{P}_\zeta(k)$. Using the constraints on the amplitude $\mathcal{A}_p$ for $\Delta=0.1$, obtained from the abundance of monochromatic PBHs in Ref.\cite{Kushwaha:2026msi}, we evaluate the present-day stochastic GW background. This is shown in Figure~\ref{fig:sigw-constraints}. As expected, the ellipsoidal collapse scenario predicts a larger GW amplitude than the spherical collapse case. Unlike the spherical approximation, it accounts for the non-spherical collapse of overdense regions~\cite{Kuhnel:2016exn,Escriva:2024lmm,Escriva:2024aeo} and therefore requires a higher collapse threshold. Consequently, for a fixed PBH abundance, a larger amplitude of primordial curvature perturbations is needed, resulting in a stronger SIGW background. To the best of our knowledge, this is the first comparison of the stochastic SIGW background predicted by the spherical and ellipsoidal PBH formation scenarios. This comparison also highlights the importance of incorporating ellipsoidal collapse in studies of PBH formation. A more detailed investigation is left for future work.

\textbf{Stochastic GW background from PBH mergers. }
After their formation, PBHs can form binaries, for example through tidal force induced by a neighboring PBH in the early Universe~\cite{2018-Sasaki.etal-CQG,Aggarwal:2025noe}. As these binaries inspiral and merge, they emit GWs that contribute to the stochastic GW background. The present-day GW energy density spectrum is given by~\cite{Aggarwal:2025noe}
\begin{align}\label{gw-merger}
    \Omega_{\rm GW,0} (f) &= \frac{f}{\rho_{c,0}} \int_0^{\frac{f_{\rm cut}}{f}-1}  
    \frac{R_{\rm PBH} (z)}{(1+z) H(z)}  \frac{d E_{\rm GW} (f_s) }{df_s} dz
\end{align}
where $\rho_{c,0} = 3 c^2 H_0^2/8\pi G $ critical energy density of the Universe today (for dimensional consistency, here we introduce speed of light $c$). The Hubble parameter at redshift $z$ is $H(z) = H_0 \sqrt{\Omega_{\gamma,0} (1+z)^4 + \Omega_{m,0} (1+z)^3 + \Omega_\Lambda}$ with Hubble constant $H_0 = 100 h \,{\rm km \, s^{-1} Mpc^{-1}}$ and $h$ is the dimensionless number which parametrizes Hubble constant. % 
The quantities $ h^2 \Omega_{m,0} = 0.142$ and $  \Omega_\Lambda = 1- \Omega_{m,0}- \Omega_{\gamma,0}$ refer to the present-day energy density fraction of non-relativistic matter, and of dark energy~\cite{Book-Durrer-2020}. 
In this work, we adopt the monochromatic PBH mass function, $\psi (m) = f_{\rm PBH} \delta (m-m_c)$, where $m_c = M_{\rm PBH}$ is the PBH mass and $f_{\rm PBH} \equiv \Omega_{\rm PBH}/\Omega_{\rm CDM}$ is the PBH fraction of cold dark matter (CDM) defined via their present-day energy density fractions.

The comoving merger rate $R_{\rm PBH}$ is given by~\cite{Raidal:2017mfl,Raidal:2018bbj,Hutsi:2020sol,Book-pbh-2025}
\begin{align}\label{r-pbh-mc}
     R_{\rm PBH} (z)  & \simeq \frac{3.13\times 10^6}{{\rm Gpc^3 \, yr}}  ~f_{\rm PBH}^{\frac{127}{37}}  \left( \frac{M_{\rm PBH}}{M_\odot}\right)^{-\frac{32}{37}} \left( \frac{t}{t_0}\right)^{-\frac{34}{37}} \, S  \,~,
\end{align}
where $t = \int_z^\infty dz'/[(1+z') H(z')]$ is the merger (cosmic) time and $t_0=13.8 \, {\rm Gyr}$ is the age of the Universe. The suppression factor $S$ (which takes $0\le S\le1$) accounts for the effects of interactions between PBH binaries and their surrounding environment in both the early and late Universe, thereby refining the merger rate relative to earlier estimates~\cite{Sasaki:2016jop,Wang:2019kaf}. 
We provide the full derivation of $R_{\rm PBH}(z)$ and the detailed analytical 
expression for $S$ in Appendix~\ref{appsec-merger-details}.

The quantity $d E_{\rm GW}/df_s$ in Eq.\eqref{gw-merger} denotes the energy spectrum of a binary in the source frame, where the source frequency $f_s$ is related to the comoving frequency as $f_s = f (1+z)$.
For non-spinning binaries, including the inspiral, merger, and ringdown phases, it is given by
\begin{align}
\frac{dE_{\rm GW}}{df_s}
=
\frac{(G\pi)^{\frac{2}{3}} \mathcal{M}_c^{\frac{5}{3}}}{3}\begin{cases}
\,f_s^{-\frac{1}{3} }\,
& f_s < f_1,
\\
w_1\,f_s^{\frac{2}{3}}\,\,
& f_1 \leq f_s < f_2,
\\%[2ex]
 \frac{w_2 \, \sigma^4 f_s^2}{( \sigma^2 + 4(f_s -f_2)^2)^2} \,
& f_2 \leq f_s < f_3 
%\\%[2ex]
%0 \,
%& f_3 \leq f_s .
\end{cases}
\end{align}
where $\mathcal{M}_c=2^{-1/5}M_{\rm PBH}$ is the chirp mass for equal-mass binaries (which is considered in this work).
The normalization constants $w_1$ and $w_2$ ensure the continuity of the spectrum. The explicit expressions for $f_{1,2,3}$ and $\sigma$ (not to be confused with the variance) are given in Appendix~\ref{appsec-energy-gw}.

Figure~\ref{fig:sgwb-typical-masses} shows the present-day stochastic GW background from PBH binary mergers for representative PBH masses, together with the sensitivities of LISA, Advanced LIGO, DECIGO, and ET. The solid (dotted) curves correspond to the suppressed (unsuppressed, $S=1$) merger-rate scenario, highlighting the significant impact of merger-rate suppression, which reduces the GW amplitude by up to a few orders of magnitude. These observations can probe PBHs in the mass range
$\in[10^{-2},1]\,M_\odot$, lighter PBHs ($M_{\rm PBH}\lesssim 10^{-2}\,M_\odot$) merge at higher frequencies (above $10\,$kHz) currently beyond their reach\footnote{Very light PBHs can also produce GWs through Hawking evaporation, with the spectrum peaking at much higher frequencies~\cite{Anantua:2008am,Dolgov:2011cq,Dong:2015yjs,Ireland:2023avg,Aggarwal:2025noe}. We do not consider this emission channel in the present work. Consequently, the relation given in Eq.\eqref{fisco-fsigw} does not apply to the evaporation-induced GW signal.}. %
Although direct detection in this high-frequency regime remains challenging, considerable effort has been devoted to indirect observational probes, theoretical constraints, and proposed laboratory-based searches for high-frequency GWs. These include radio and CMB observations, as well as a variety of experimental concepts~\cite{2006-Cruise-CQG,2008-Nishizawa.etal-PRD,2012-Cruise-CQG,2017-Chou.etal-PRD,2021-Domcke.Garcia-Cely-PRL,Domcke:2022rgu,Bringmann:2023gba,2022-Kushwaha.etal-MNRAS,2023-Kushwaha.Sunil.Shanki-IJMPD,Ito:2023nkq,Ito:2023fcr,He:2023xoh,Domcke:2024mfu,Gatti:2024mde,Cai:2025fpe,Kushwaha:2025mia,Pappas:2025zld,Kim:2025izt,Aggarwal:2025noe,Matsuo:2026lvo,Gorji:2026tln}.
\begin{figure}[t!]
\centering
%\hspace{.2cm}
\includegraphics[height=2.4in,width=3.7in]{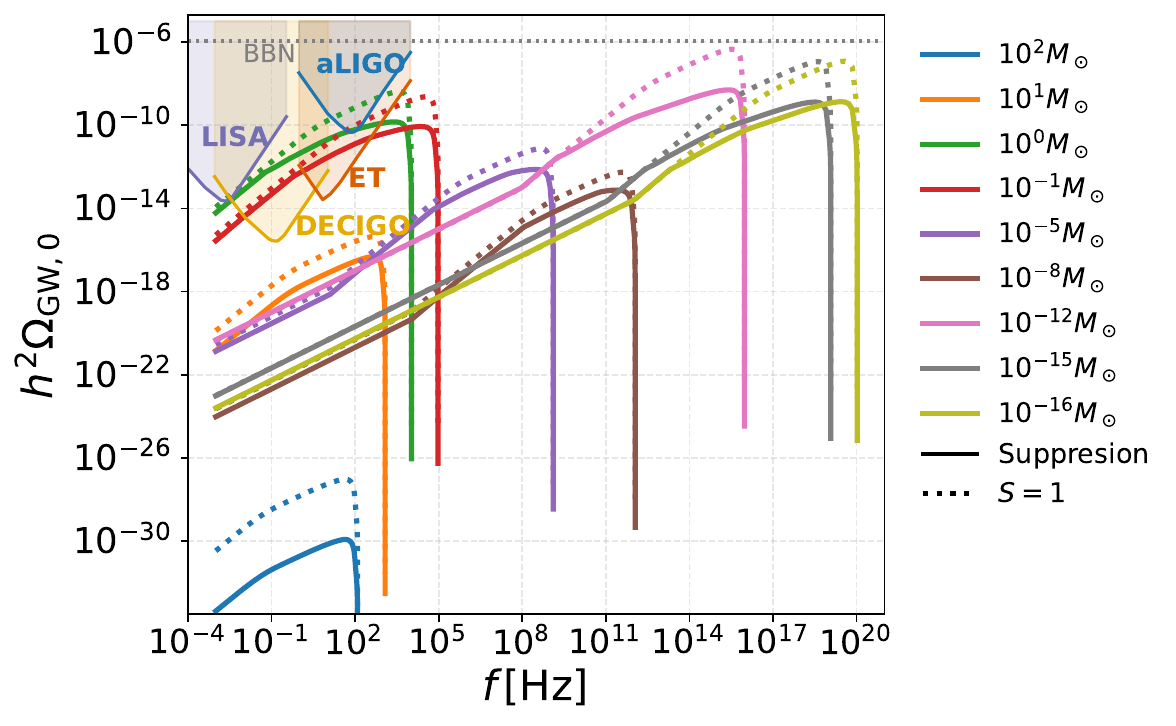}
\caption{Stochastic GW background from PBH binary mergers for representative PBH masses in the range $10^{-16}$--$10^{2}\,M_\odot$. The spectra are computed using Eq.~\eqref{gw-merger} for the corresponding $f_{\rm PBH}$ values from Ref.~\cite{2021-Carr.etal-Rept.Prog.Phys}, with merger-rate suppression (solid curves) and without suppression ($S=1$; dotted curves). Sensitivity curves are the same as in Figure~\ref{fig:sigw-constraints}.
}
\label{fig:sgwb-typical-masses}
\end{figure}
\begin{figure}[t!]
\centering
%\hspace{.2cm}
\includegraphics[height=2.3in,width=3.6in]{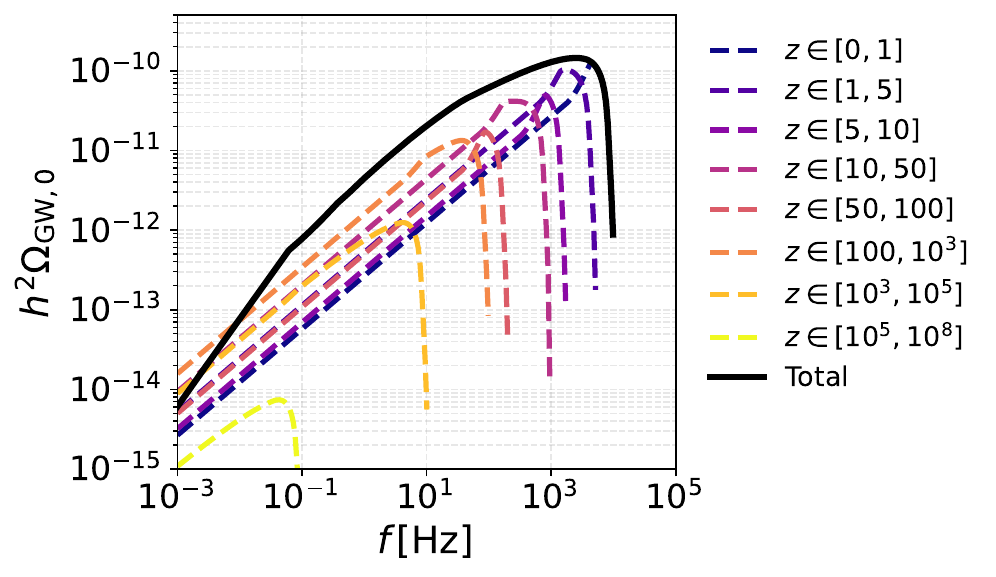}
\caption{Redshift-bin contributions (dashed curves) to the stochastic GW background from PBH binary mergers for a representative PBH mass of $1\,M_\odot$, corresponding to the suppressed spectrum shown in Fig.~\ref{fig:sgwb-typical-masses}. The black solid curve denotes the total integrated estimate.} 
\label{fig:sgwb-redshift-dominant}
\end{figure}
An interesting implication of the combined PBH formation and merger scenario is that it establishes a direct connection between the low-frequency stochastic SIGW background generated from the same enhanced primordial curvature perturbations that produce PBHs and the high-frequency GW signal emitted by subsequent binary mergers. This correspondence is determined entirely by the PBH mass. %

For equal-mass binaries, the inspiral-to-merger transition occurs near the ISCO frequency, $f_{\rm ISCO} = 2200~{\rm Hz}\,(M_\odot/M_{\rm PBH})$. Using the PBH mass relation in Eq.~\eqref{mpbh-k-relation} (neglecting the $g_*$), % 
we obtain
\begin{align}\label{fisco-fsigw}
f_{\rm ISCO}
\simeq
3.4\times10^{20}\ {\rm Hz}
\left(\frac{f_{\rm SIGW}}{{\rm Hz}}\right)^2.
\end{align}
This directly connects the frequency of stochastic SIGW background to the characteristic binary merger frequency. %
To verify that $f_{\rm ISCO}$ in Eq.~\eqref{fisco-fsigw} corresponds to the present-day observed frequency, Fig.~\ref{fig:sgwb-redshift-dominant} shows that the merger background is dominated by low-redshift contributions. Thus, $f_{\rm ISCO}$ represents to the present-day observed frequency.
\begin{figure}[t!]
\centering
%\hspace{.2cm}
\includegraphics[height=2.8in,width=3.4in]{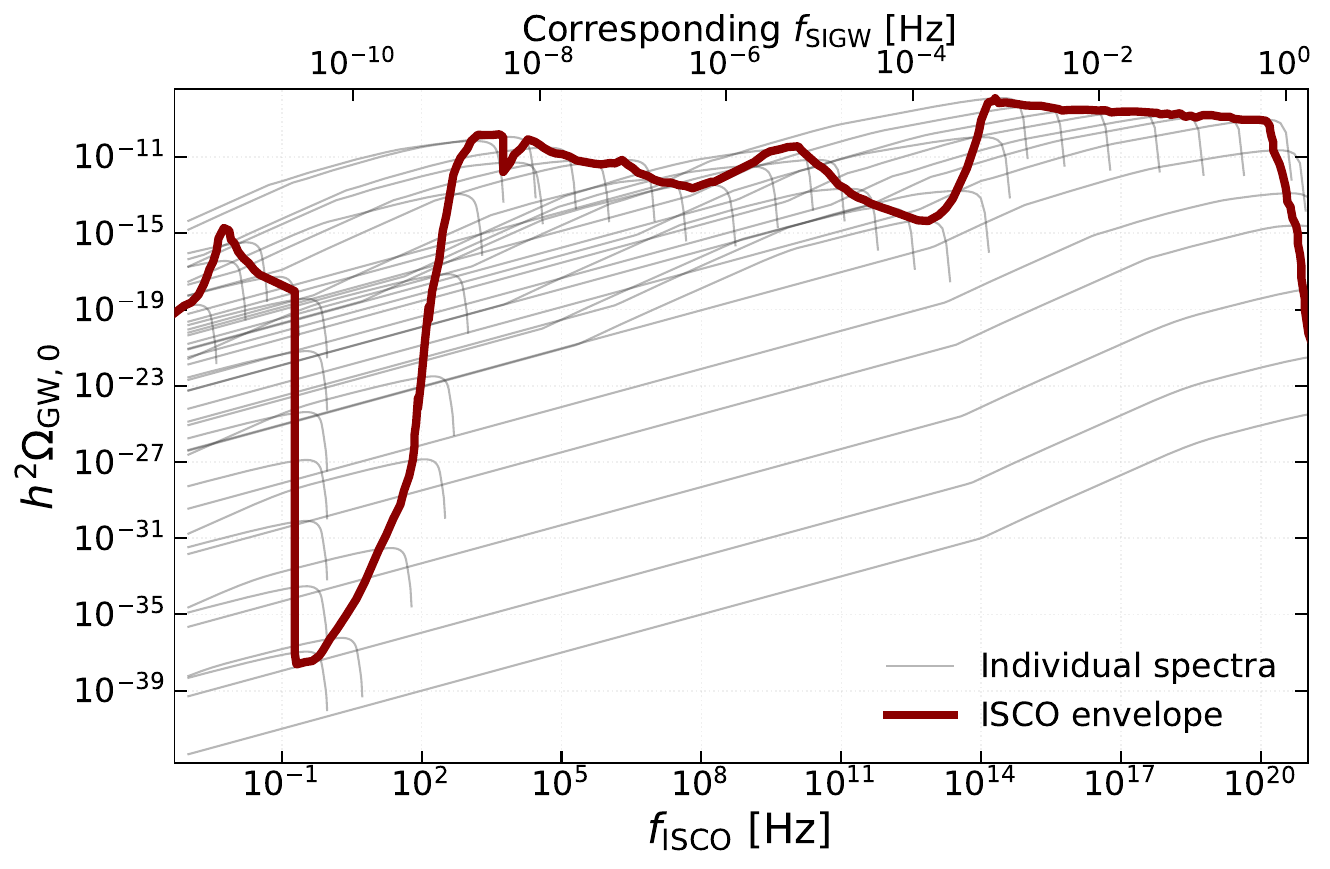}
\caption{Stochastic GW background from PBH binary mergers, shown as the dark-red curve (envelope), as a function of the ISCO frequency,
$f_{\rm ISCO}=2200\,{\rm Hz}\,(M_\odot/M_{\rm PBH})$.
The individual spectra from Fig.~\ref{fig:sgwb-typical-masses} are shown as gray curves for 50 representative PBH masses.
The top axis, shown for reference, gives the corresponding SIGW frequency from Eq.~\eqref{fisco-fsigw}. The estimates are for the suppressed merger-rate.%
}
\label{fig:sgwb-envelop}
\end{figure}
Figure~\ref{fig:sgwb-envelop} illustrates the correspondence between the merger and SIGW frequencies through the relation~\eqref{fisco-fsigw}. The gray curves show the merger-induced stochastic GW background for 50 representative PBH masses, while the dark-red envelope traces the corresponding amplitudes evaluated at $f_{\rm ISCO}$. The upper horizontal axis is shows the corresponding SIGW frequency using Eq.~\eqref{fisco-fsigw}. As evident from Fig.~\ref{fig:sgwb-envelop}, the peaks of the merger spectra (at frequency $f_{\rm peak}$) do not coincide with the $f_{\rm ISCO}$ envelope. This is expected since General Relativity predicts $f_{\rm peak}=\mathcal{O}(1)\,f_{\rm ISCO}$~\cite{Book-Maggiore-vol-1,Book-Maggiore-Vol2}. Numerically, we obtain the nearly mass-independent relation $f_{\rm peak}\simeq1.79\,f_{\rm ISCO}$ for the unsuppressed case ($S=1$) (see \ref{appfig:sgwb-envelop} in Appendix~\ref{appsec-unsuppress}), while the suppressed case shows a weak mass dependence in the range $1 \lesssim f_{\rm peak}/f_{\rm ISCO}\lesssim 4$.
Finally, we emphasize that the one-to-one correspondence in Eq.~\eqref{fisco-fsigw} is specific to monochromatic PBHs and breaks down for extended PBH mass functions arising from broad primordial power spectra.

\textbf{Conclusion.}
We established a direct connection between low-frequency SIGWs and the high-frequency stochastic background from PBH binary mergers, assuming a shared origin in enhanced primordial curvature perturbations. Using PBH abundance constraints, we computed the corresponding SIGW background, demonstrating that the ellipsoidal collapse yields a stronger signal than the spherical approximation. We also computed the corresponding PBH merger background, explicitly quantifying the impact of merger-rate suppression compared to idealized unsuppressed scenarios. We derived a model-independent mapping between the peak SIGW frequency and the characteristic merger ISCO frequency. 
This unified framework enables complementary probes of the same small-scale primordial fluctuations across widely separated frequency bands, allowing existing and forthcoming low-frequency GW observations to indirectly constrain otherwise inaccessible high-frequency PBH merger signals. The formalism can be extended to broad primordial curvature power spectra and the resulting extended PBH mass functions.

\textbf{Acknowledgements.}
The work of A.K. was supported by the Japan Society for the Promotion of Science (JSPS) as part of the JSPS Postdoctoral Program (Grant Number: 25KF0107). A.K. is grateful to Teruaki Suyama for valuable discussions and insightful comments on the manuscript. A.K. also thanks S. Shankaranarayanan for helpful comments on the manuscript.

\newpage
\appendix
\onecolumngrid

\section{Details of the PBH merger rate}
\label{appsec-merger-details}

In this appendix, we summarize the expressions used to compute the PBH merger rate and the suppression factor introduced in the main text. Unless otherwise stated, we follow the same notation.

The comoving merger rate $R_{\rm PBH}$ at redshift $z$ is given by~\cite{Raidal:2017mfl,Raidal:2018bbj,Hutsi:2020sol,Book-pbh-2025} 
\begin{align}\label{r-pbh-full}
     dR_{\rm PBH} (z) 
      \simeq \frac{1.6\times 10^6}{{\rm Gpc^3 \, yr}} \int d m_1 \int dm_2 ~f_{\rm PBH}^{\frac{53}{37}} \left( \frac{m_1 m_2}{(m_1+m_2)^2}\right)^{-\frac{34}{37}} 
    \left( \frac{M}{M_\odot}\right)^{-\frac{32}{37}} \left( \frac{t}{t_0}\right)^{-\frac{34}{37}} \, S (m,\psi,f_{\rm PBH}) \,\, \psi(m_1) \psi (m_2)~.
\end{align}
The PBHs mass function $\psi (m)$ is normalized as $f_{\rm PBH} = \int_0^\infty \psi (m) dm$, where the abundance of PBH in cold dark matter (CDM) is defined by the dimensionless parameter $f_{\rm PBH} \equiv \Omega_{\rm PBH}/\Omega_{\rm CDM}$, with $\Omega_{\rm PBH}$ and $\Omega_{\rm CDM}$ representing the present day energy density fractions of PBH and CDM. 

The suppression factor $S (m,\psi,f_{\rm PBH})$, which satisfies $0\le S\le1$, accounts for the effects of interactions between PBH binaries and their surrounding environment in both the early and late Universe, thereby refining the merger rates of Refs.~\cite{Sasaki:2016jop,Wang:2019kaf}. An analytic expression for $S(m,\psi,f_{\rm PBH})$, supported by numerical simulations, has been derived in Refs.~\cite{Raidal:2017mfl,Raidal:2018bbj,Hutsi:2020sol,Book-pbh-2025} and is given by
\begin{align}
    S(m,\psi,f_{\rm PBH}) =  S_L \frac{e^{- \bar{N}(y)}}{\Gamma (21/37)} \int dv v^{-\frac{16}{37}} \exp \left[ - \bar{N}(y) \langle m \rangle  \int \frac{dm}{m} \psi (m) F\left( \frac{m}{\langle m \rangle} \frac{v}{\bar{N}(y)}\right) - \frac{3 \sigma_M^2 v^2}{10 f_{\rm PBH}^2} \right]
\end{align}
where $\sigma_M^2=0.006$ denotes the rescaled variance of matter density perturbations at binary formation~\cite{Hall:2020daa,Raidal:2018bbj}, $\langle m\rangle$ is the mean PBH mass, and $\bar{N}(y)$ is the expected number of PBHs within the comoving radius $y$ surrounding the binary given by
\begin{equation}\label{nbar}
    \bar N(y)\simeq
\frac{M_t}{\langle m\rangle}
\frac{f_{\rm PBH}}{f_{\rm PBH}+\sigma_M}~~.
\end{equation}
The function $F(x)$ is expressed in terms of the generalized hypergeometric function~\cite{Raidal:2018bbj,Hutsi:2020sol,Hall:2020daa} as
\begin{align}
  F(x)\equiv
{}_1F_2\!\left(-\frac12;\frac34,\frac54;-\frac{9x^2}{16}\right)-1  ~~.
\end{align}
The late-time suppression factor $S_L$ is approximated by
 \begin{align}
    S_L(t,f_{\rm PBH}) \approx \min \left\{1,0.01\left[ (t/t_0)^{0.44} f_{\rm PBH} \right]^{-0.65} e^{0.03 \, \ln^2 [(t/t_0)^{0.44} f_{\rm PBH} ]} \right\}~.
\end{align}   
\subsection{Monochromatic case}
For the monochromatic mass function, $\psi(m)=f_{\rm PBH}\delta(m-m_c)$, we obtain $\langle m\rangle = m_c $ and $\bar{N}(y) = \frac{2f_{\rm PBH}}{(f_{\rm PBH}+\sigma_M)}$. Throughout this work, we identify $m_c=M_{\rm PBH}$. The suppression factor can be calculated as
\begin{align}
    S (M_{\rm PBH},f_{\rm PBH}) =  S_L \frac{e^{- \bar{N}(y)}}{\Gamma (21/37)} \int dv v^{-\frac{16}{37}} \exp \left[ - \frac{2 \, f_{\rm PBH}^2}{(f_{\rm PBH} + \sigma_M)} F\left( \frac{M_{\rm PBH } \, v \, (f_{\rm PBH} + \sigma_M)}{2}\right) - \frac{3 \sigma_M^2 v^2}{10 f_{\rm PBH}^2} \right]
\end{align}

Lastly, we would like to provide the expression for the case of no suppression, which is given by

\begin{align}\label{ap-r-pbh-mc}
     R_{\rm PBH} (z)  & \simeq \frac{3.13\times 10^6}{{\rm Gpc^3 \, yr}}  ~f_{\rm PBH}^{\frac{127}{37}}  \left( \frac{m_c}{M_\odot}\right)^{-\frac{32}{37}} \left( \frac{t}{t_0}\right)^{-\frac{34}{37}} \, \,~.
\end{align}

\section{Details on Energy spectrum of GWs}
\label{appsec-energy-gw}
Following Refs.~\cite{Wang:2019kaf}, we summarize the expressions and numerical values used to evaluate the GW energy spectrum, $dE_{\rm GW}/df_s$, appearing in the main text.
\begin{align}
\frac{dE_{\rm GW} (f_s)}{df_s}
=
\frac{(G\pi)^{\frac{2}{3}} \mathcal{M}_c^{\frac{5}{3}}}{3}\begin{cases}
\,f_s^{-\frac{1}{3} }\,
& f_s < f_1,
\\
w_1\,f_s^{\frac{2}{3}}\,\,
& f_1 \leq f_s < f_2,
\\%[2ex]
 \frac{w_2 \, \sigma^4 f_s^2}{( \sigma^2 + 4(f_s -f_2)^2)^2} \,
& f_2 \leq f_s < f_3 ~,
\\%[2ex]
0 \,
& f_3 \leq f_s .
\end{cases}
\end{align}
where $\mathcal{M}_c^{5/3}=m_{1}m_{2}(m_{1}+m_{2})^{-1/3}$ is the chirp mass, $f_s$ is the GW frequency in the source frame. The normalization constants $w_1,w_2$ are chosen to ensure the continuity of the spectrum. The other parameters, such as $f_{1,2,3}$ and $\sigma$ (here it should not be confused with variance) can be expressed in terms of total mass parameter $M_t=m_1+m_2$ and symmetric mass ratio $\tau = m_1 m_2/(m_1+m_2)^2$. For the equal-mass binaries considered in this work, where $m_1=m_2 = M_{\rm PBH}$, gives $\tau = 0.25, \,M_t = 2 M_{\rm PBH}$ and $\mathcal{M}_{c}= 2^{-1/5} \, M_{\rm PBH}$. Therefore, we obtain~\cite{Wang:2019kaf,Ajith:2009bn}
\begin{align*}
\pi M_{t} f_1 &=(1-4.455+3.521)+0.6437\tau-0.05822\tau^2-7.092\tau^3 \quad \implies f_1 = 0.0358015 \,M_t^{-1} = 0.0179 \, M_{PBH}^{-1} \\
\pi M_{t} f_2 &= (1-0.63)/2+0.1469\tau-0.0249\tau^2+2.325\tau^3 \quad \implies f_2 = 0.081645 \,M_t^{-1} = 0.0408 \, M_{PBH}^{-1} \\
\pi M_{t} \sigma &= (1-0.63)/4 -0.4098\tau +1.829\tau^2-2.87\tau^3 \quad \implies \sigma = 0.0189454 \,M_t^{-1} = 0.0094 \, M_{PBH}^{-1} \\
\pi M_{t} f_3 &= 0.3236 -0.1331\tau -0.2714\tau^2 +4.922\tau^3 \quad \implies f_3 = 0.111494 \,M_t^{-1} = 0.0557\, M_{PBH}^{-1}
\end{align*}
The cutoff frequency is determined by $f_3$ as
$f_{\rm cut}=f_3=0.0557\,M_{\rm PBH}^{-1}$.
The fitting formulae of Ref.~\cite{Ajith:2009bn} are expressed in geometrized units ($G=c=1$). Therefore, care must be taken when converting the characteristic frequencies to physical units. Using the standard relation $1\,M_\odot = 1.477~{\rm km} \simeq 4.92\times10^{-6}~{\rm s}$, the cutoff frequency can be written as
\begin{align}
f_{\rm cut}
=
0.0557
\left(\frac{M_\odot}{M_{\rm PBH}}\right)
\frac{1}{M_\odot}
\simeq
11~{\rm kHz}
\left(\frac{M_\odot}{M_{\rm PBH}}\right),
\end{align}
which is more convenient for practical applications. Similarly, the remaining characteristic frequencies are
\begin{align*}
 f_1  \simeq 3.6~{\rm kHz}\left(\frac{M_\odot}{M_{\rm PBH}}\right), ~
 f_2  \simeq 8.2~{\rm kHz}\left(\frac{M_\odot}{M_{\rm PBH}}\right), ~
 \sigma \simeq 2.0~{\rm kHz}\left(\frac{M_\odot}{M_{\rm PBH}}\right),~
 f_3  \simeq 11~{\rm kHz}\left(\frac{M_\odot}{M_{\rm PBH}}\right).
\end{align*}
For illustration, a PBH with mass $M_{\rm PBH}=10^{-18}M_\odot$ corresponds to $f_{\rm cut}\simeq10^{22}~{\rm Hz}$.

Finally, the normalization constants $w_1$ and $w_2$ are determined by requiring the continuity of the spectrum at the transition frequencies. Continuity at $f_1$ gives
\begin{align}
w_1=\frac{1}{f_1},
\end{align}
while continuity at $f_2$ implies
\begin{align}
w_1f_2^{2/3}=w_2f_2^2, \qquad \implies \quad w_2=\frac{1}{f_1f_2^{4/3}} ~~.
\end{align}
\begin{figure}[t!]
\centering
%\hspace{.2cm}
\includegraphics[height=3in,width=5in]{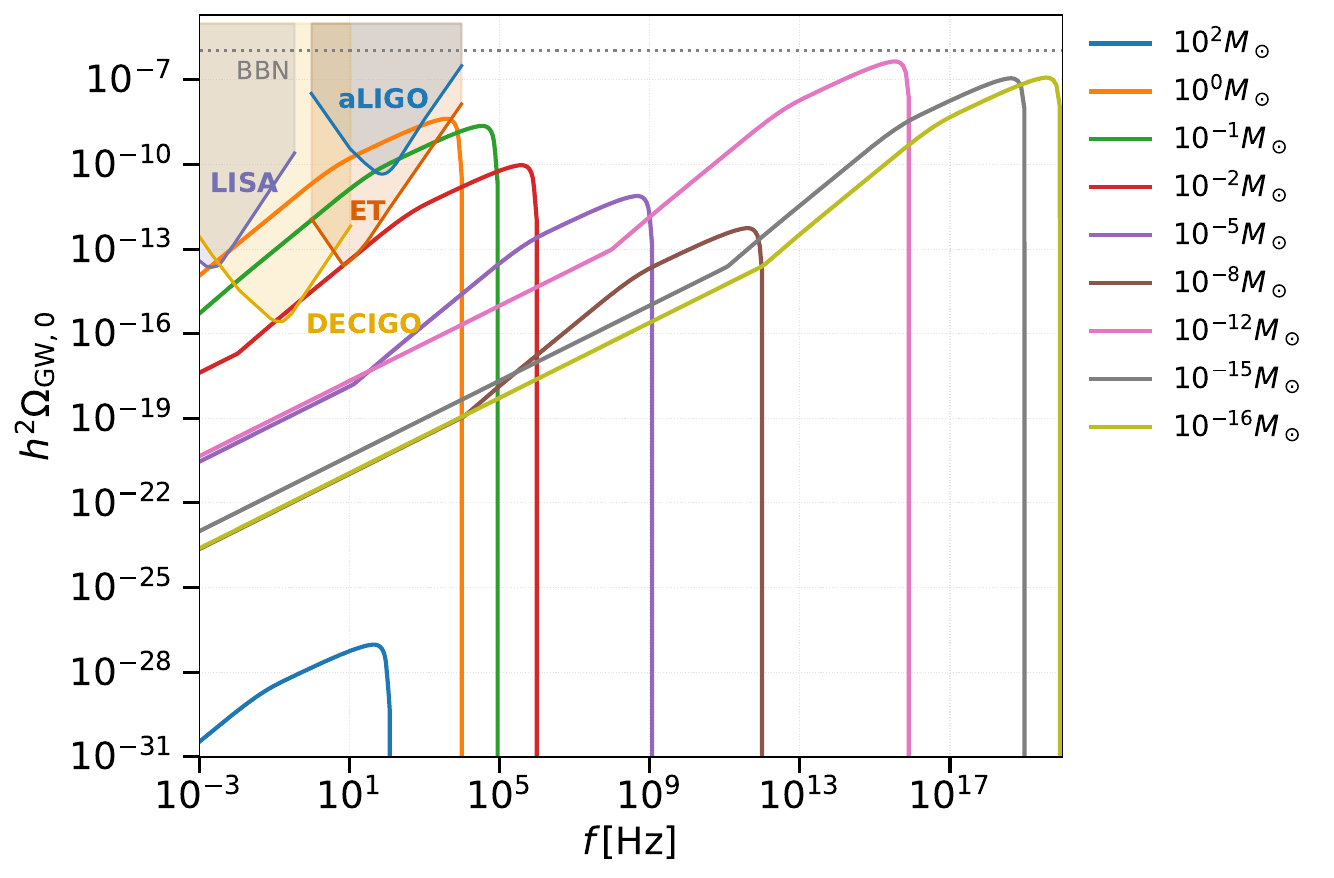}
\caption{Stochastic GW background produced by mergers of PBH binaries for representative PBH masses spanning $10^{-16} - 10^{2}M_\odot$. For each mass, the PBH abundance, $f_{\rm PBH}$, is taken from Ref.\cite{2021-Carr.etal-Rept.Prog.Phys}. The spectra are computed using Eq.(6) in the main text for $S=1$. Sensitivity curves and the BBN bound are the same as in the main text for Figure~1.
}
\label{apfig:sgwb-typical-masses}
\end{figure}
\begin{figure}
\centering
%\hspace{.2cm}
\includegraphics[height=4in,width=6in]{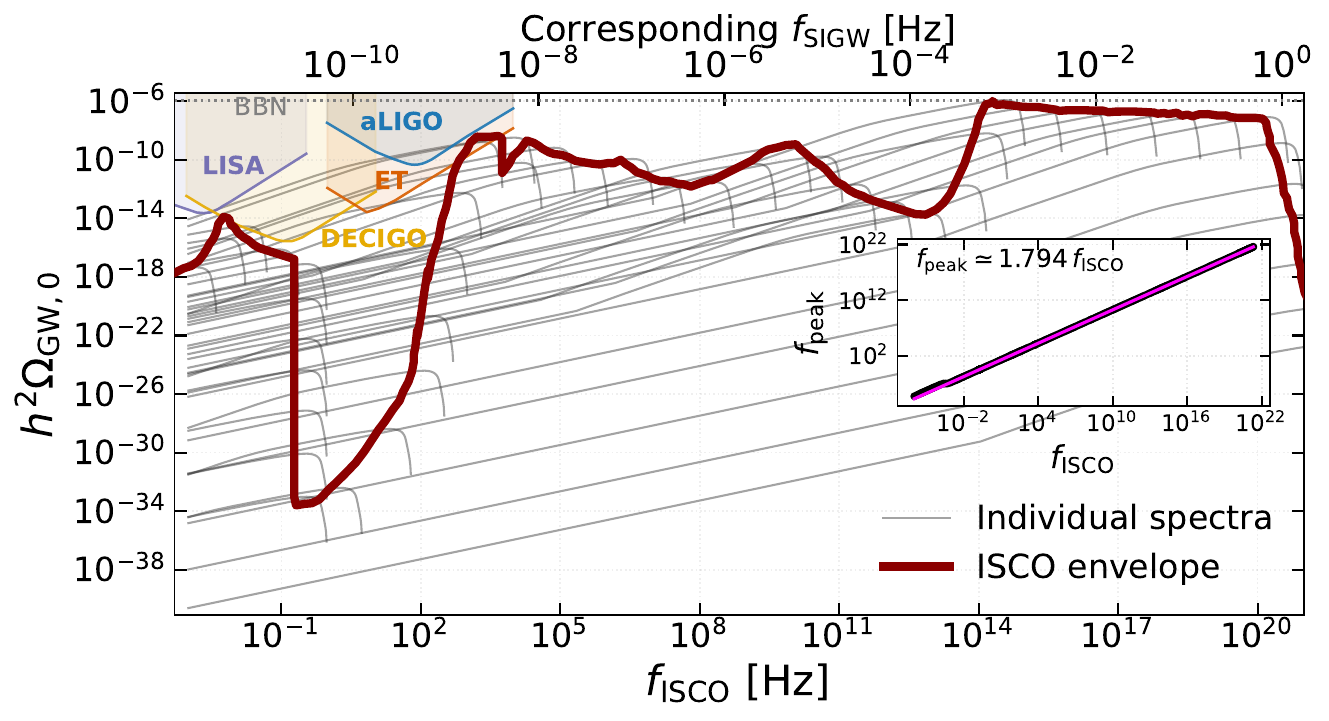}[h!]
\caption{Stochastic GW background from PBH binary mergers, shown as the dark-red curve (envelope), as a function of the ISCO frequency,
$f_{\rm ISCO}=2200\,{\rm Hz}\,(M_\odot/M_{\rm PBH})$.
The individual spectra from \ref{apfig:sgwb-typical-masses} are shown as gray curves for 50 representative PBH masses.
The top axis, shown for reference, gives the corresponding SIGW frequency obtained using Eq.(9) in the main text.
The inset shows that the numerical peak frequency of the individual spectrum (gray curves) follows the nearly mass-independent relation
$f_{\rm peak}\simeq1.79\,f_{\rm ISCO}$, with the empirical fit shown in magenta. As in Figure~1 in the main text, the sensitivity curves and the BBN bound are plotted directly against the horizontal frequency, $f_{\rm ISCO}$.
}
\label{appfig:sgwb-envelop}
\end{figure}
\section{Plots of $h^2 \Omega_{\rm GW,0}$ for the unsuppressed case}
\label{appsec-unsuppress}
In this section, we provide the plots for the standard case of no suppression ($S=1$).
These are shown in \ref{apfig:sgwb-typical-masses} and \ref{appfig:sgwb-envelop} corresponding to figures (2) and (4) in the main text, respectively.

\bibliography{References}
\end{document}